\documentclass[sigconf]{acmart}
\usepackage{graphicx}
\usepackage{amsmath}
\usepackage{booktabs}
\usepackage{algorithm}
\usepackage{algorithmic}
\usepackage{amsfonts}
\usepackage{multirow}
\usepackage{makecell}
\usepackage{subfigure}
\usepackage{color}
\usepackage{bm}
\usepackage{epstopdf}
\usepackage{url}
\usepackage[cal=cm]{mathalfa}
\usepackage{balance}
\usepackage{threeparttable}
\usepackage{lipsum}
\usepackage{enumitem}

\AtBeginDocument{%
  \providecommand\BibTeX{{%
    \normalfont B\kern-0.5em{\scshape i\kern-0.25em b}\kern-0.8em\TeX}}}

\begin{document}
\title{Moment\&Cross: Next-Generation Real-Time Cross-Domain CTR Prediction for Live-Streaming Recommendation at Kuaishou}
\renewcommand{\shorttitle}{Moment\&Cross}

 \author{Jiangxia Cao, Shen Wang, Yue Li, Shenghui Wang, Jian Tang, Shiyao Wang, \\Shuang Yang, Zhaojie Liu, Guorui Zhou}
\affiliation{
  \institution{Kuaishou Technology, Beijing, China}
  \country{\{caojiangxia, wangshen, liyue16, wangshenghui03, tangjian03, wangshiyao08, \\yangshuang08, zhaotianxing, zhouguorui\}@kuaishou.com}
 }

\begin{abstract}
Kuaishou, is one of the largest short-video and live-streaming platform, compared with short-video recommendations, live-streaming recommendation is more complex because of: (1) temporarily-alive to distribution, (2) user may watch for a long time with feedback delay, (3) content is unpredictable and changes over time.
Actually, even if a user is interested in the live-streaming author, it still may be an negative watching (e.g., short-view < 3s) since the real-time content is not attractive enough.
Therefore, for live-streaming recommendation, there exists a challenging task: \textit{how do we recommend the live-streaming at right moment for users?}
Additionally, our platform's major exposure content is short short-video, and the amount of exposed short-video is 9x more than exposed live-streaming. 
Thus users will leave more behaviors on short-videos, which leads to a serious data imbalance problem making the live-streaming data could not fully reflect user interests.
In such case, there raises another challenging task: \textit{how do we utilize users' short-video behaviors to make live-streaming recommendation better?}
For the first challenge, we analyzed our data and observed an interesting phenomenon: when a live-streaming is at a high-light moment (e.g., Dancing), its click rate (CTR) will be increasing significantly.
Inspired by our observation, we believe there exists a potential solution to automatically identify which live-streaming may be at its `high-light moment', based on the wisdom of crowds of many user behaviors towards the live-streaming current moment.
Therefore, our goal is to enable model to perceive all behaviors occurring as soon as possible, so that model can know which live-streaming is in the CTR increasing status.
To achieve the idea, we have upgraded our data-streaming engine to real-time 30s report manner and devised a novel first-only mask learning strategy to supervise our model, named \textbf{Moment}.
For the second challenge, we mainly follow the search-based interest modeling idea: first devise General Search Units (GSUs) to search users' short-video/live-streaming history, and then use Extract Search Units (ESUs) to compress them.
Besides, we also introduce a contrastive objective to align short-videos and live-streaming embedding spaces to enhance their correlation, named \textbf{Cross}.
We conduct extensive offline/online and ablation studies to verify our Moment\&Cross effectiveness.
\end{abstract}



\maketitle

\section{Introduction}
\begin{figure}[th!]
\begin{center}
\includegraphics[width=8cm,height=10cm]{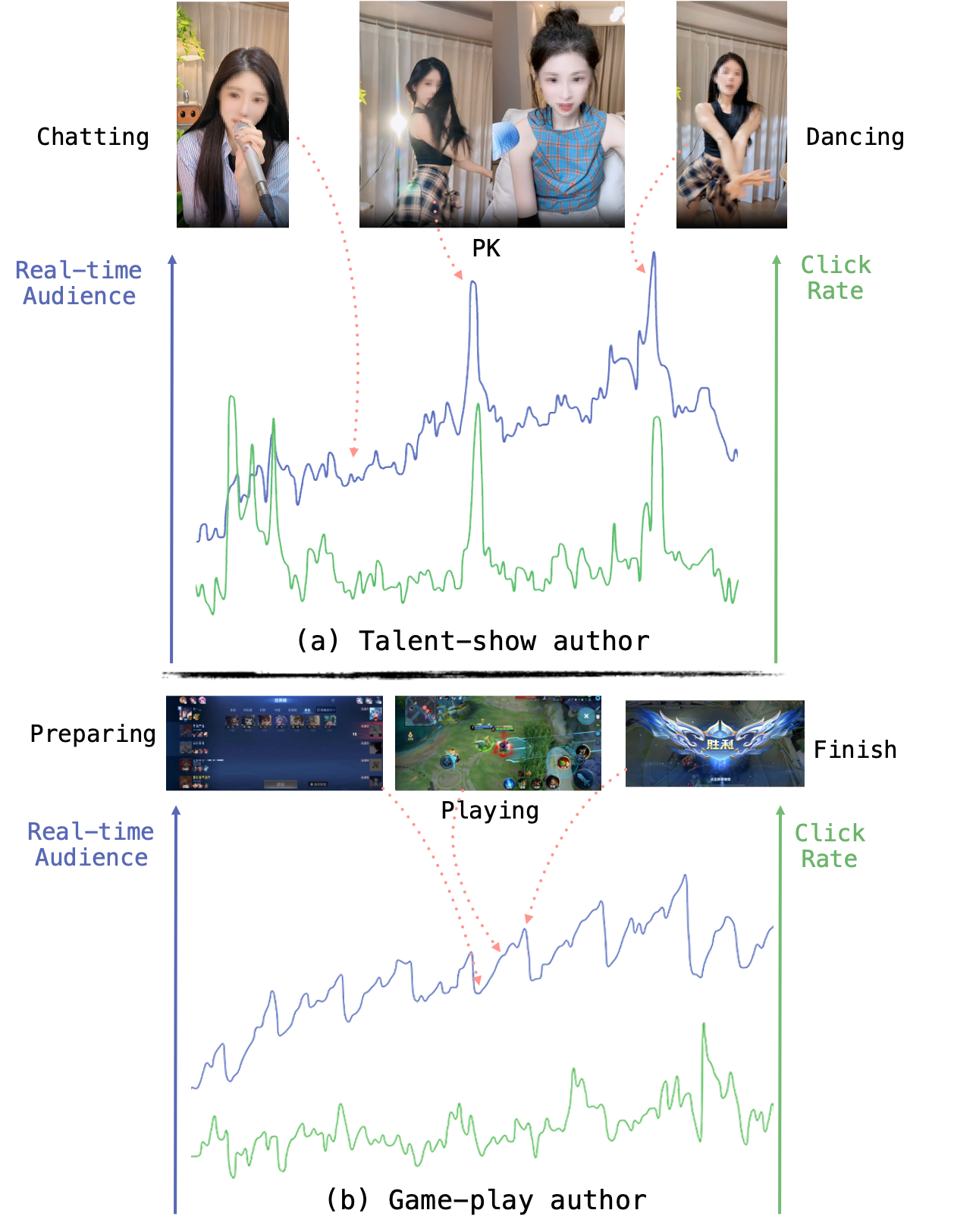}
\caption{Typical live-streaming pattern between the `high-light moment', real-time audience and the CTR trends.}
\label{crossmomentpbservation}
\end{center}
\end{figure}

\begin{figure*}[t]
\begin{center}
\includegraphics[width=14cm,height=8cm]{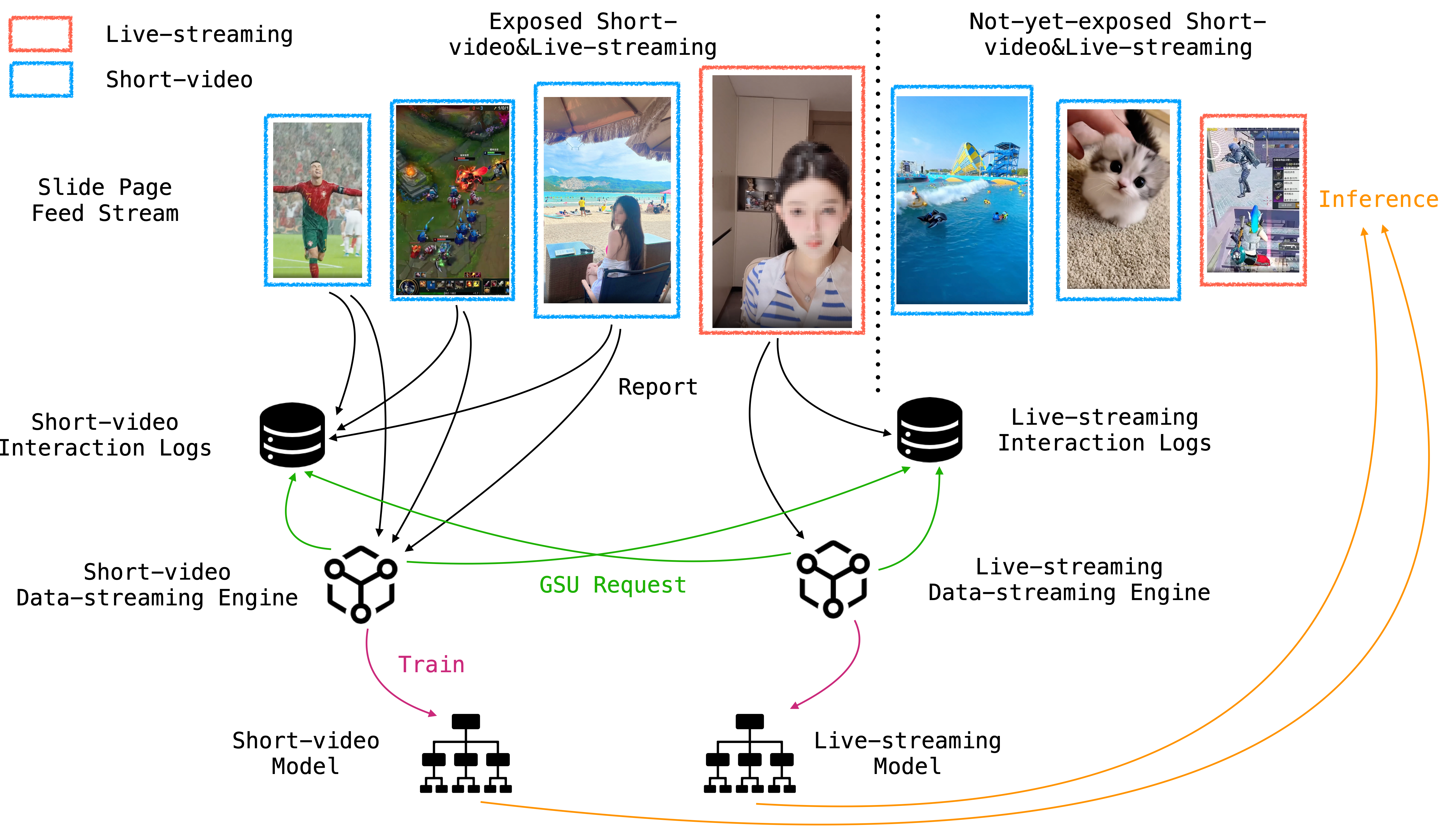}
\caption{The Slide page RecSys architecture of Kuaishou short-video and live-streaming services, different services are separately with their own data-streaming and model. The only way to access users' other business logs, is by utilizing the `interaction logs' storage services to retrospect historical user's short-video behaviors to find related small group items.
}
\label{recarchitecture}
\end{center}
\end{figure*}

Short-video and live-streaming platform like Kuaishou and Douyin have grown rapidly in recent years, attracting a lot of attention and accumulating a large number of active users.
At Kuaishou, users always watch content at the \textbf{slide} page: the content on this page will automatically be played according to the user’s up and down scrolling on screen.
Therefore, a powerful RecSys~\cite{pepnet, twin} is the foundation of our service, to influence user watching experience to decide what content to watch next for users.
Compared with the wide-explored short-video recommendation~\cite{,trinity,interestclock,youtubemmoe,counterfactual}, live-streaming recommendation~\cite{liverec} is a harder task since different natures of media as: 
(1) \textit{Temporarily life-cycle}: Different from short-video has an eternal life-cycle to distribute, live-streaming is temporary (average 1 hour) to distribute in our system. 
(2) \textit{Long-term feedback delay~\cite{handlingdelay}}: Unlike short-videos, their average duration is about 55s, thus the report manner could quickly report all user behaviors to supervise model training.
Nevertheless, the live-streaming is much longer, and some sparse and \textbf{valuable feedback may happen after watching half an hour}, e.g., users buy digital gifts for the live-streaming author.
(3) \textit{Dynamic content change~\cite{xi2023multimodal}}: Different from short videos always playing from 0s, live-streaming content is ever-changing, thus for a live-streaming, a user watching at \textbf{different time points maybe have distinct behaviors}.

Therefore, our live-streaming RecSys needs to solve a very challenging problem: \textit{how do we recommend the live-streamings at right moment for users?}
To answer the question, we show two live-streaming author cases in Figure~\ref{crossmomentpbservation}:
(1) For the talent-show live-streaming: their authors spend a lot of time chatting with audiences, participating PK with other authors, and showing their talents sometimes. In terms of user behaviors, there is a significant difference between high-light moments and chatting: when the author shows dancing talent, the number of users entering live-streaming will increase significantly, and the user will exit after finish talent showing.
(2) For the game-play live-streaming: their authors play game competitions one by one. During the author playing at one competition, the live-streaming real-time audience will continue to accumulate until the end of the competition. At the competition finished, the number of watching audience dropped significantly.

Actually, no matter what category the author is, users always tend to click the live-streaming for better watching live-streaming high-light moments, but the live-streaming content is ever-changing, so it is not an easy task to identify which live-streaming will show highlight moment.
Fortunately, there may exist a potential solution to automatically identify which live-streaming is at its `high-light moment', based on the wisdom of many user behaviors towards the live-streaming current moment: in Figure~\ref{crossmomentpbservation}, the CTR trend shows highly consistency relation with user click behaviors (i.e., peaked and troughed at the same time).
Therefore, if our model could \textbf{capture the increasing CTR trend}, then the model can discover potential highlight moments based on a large amount of user positive feedback.

Aside from the highlight moments capturing challenge, our live-streaming model has a more serious problem: data sparsity.
On the slide page, users can watch short-videos and live-streamings in an interspersed manner according to the user’s up-and-down scrolling on screen.
Nevertheless, the slide page is about 90\% of the exposed content is short-videos, thus our live-streaming RecSys has the risk could not fully learning the user interests to make precise CTR predictions.
In such case, there raises another challenging task: \textit{how do we utilize users' rich short-video behaviors to make live-streaming recommendation better?}
To answer the question, we first explain the workflow of our architecture (as shown in Figure~\ref{recarchitecture}).
In industry, the different businesses are deployed separately, for instance, the users' short-video real-time interactions (e.g., long-view, like and so on) are only assembled by the short-video data-streaming engine to be organized into specific training sample data formats.
Then, the short-video model can consume short-video data-streaming to fit real-time data distribution to make precise recommendation.
Since the different data-streaming engines will generate different data formats training samples, thus our live-streaming model can only be supervised by the users' live-streaming data-streaming.
Although we cannot consume short-video data-streaming directly, fortunately, we have constructed historical storage services to save users' interaction logs~\cite{eta,sdim}, and the data-streaming engine could send requests to obtain users' interaction history from other businesses to \textbf{assemble them as a part of input features}.
Therefore, we can align the live-streaming and short-video embedding space, then our model could utilize users short-video interests to determine which live-streaming with a similar style should be recommended.

In this paper, we present our effective and efficient solutions \textbf{Moment\&Cross} - towards building the next-generation live streaming framework.
For the first challenge, our goal is to encourage live-streaming model to be able to perceive what kind of live-streaming has the \textbf{CTR increasing trends}.
Therefore, we need to utilize the real-time occurring user behaviors to train our model as soon as possible, to capture the real-time CTR status for each live-streaming.
As shown in Figure~\ref{recarchitecture}, the CTR signal is first reported to our live-streaming data-streaming engine and then fed to our model.
However, as many industrial RecSys implemented, the report module needs to wait for a while (e.g., 5 minutes) to collect enough behaviors and then report them to the data-streaming engine at once.
Particularly, our live-streaming services also employ a fixed-window 5-minute style data-streaming to support our model for several years, however, a 5-minute feedback delay is not real-time enough to support our model to capture the CTR increasing trend.
To this end, we have upgraded our training framework from Fast-Slow report manner to real-time 30s report manner, and devised a novel first-only mask learning strategy to supervise our model, named \textbf{Moment}.
For the second challenge, our goal is to \textbf{exploit users historical short-video sequences} and align their embedding space with live-streaming.
In fact, the user's short-video history is too long to model directly~\cite{twinv2} (e.g., a high-active user easily watch 10,000 short-videos in 1 month),  thus we mainly follow the cascading search-based~\cite{sim} interest modeling framework: (1) introduce General Search Units (GSUs) to retrospect user life-long history and then filter to obtain a top related item sequence (hundreds-level); (2) devise Exact Search Units (ESUs) to compress sequence information to obtain user interests, e.g, sequence pooling~\cite{youtube}, target-item-attention~\cite{din}.
Besides, we also introduce a contrastive~\cite{mine} objective to align short-videos and live-streaming embedding spaces to enhance their correlation, thus our model could distribute live-streaming with a similar style according to users' interests from rich short-video interaction history, named \textbf{Cross}.
In summary, our contributions are as follows:
\begin{itemize}[leftmargin=*,align=left]
    \item We present a novel real-time learning framework to discover the `high-light' live-streaming automatically, towards building next-generation live-streaming recommendation.
    \item We devise simple-yet-effective techniques to transfer user short-video interests for live-streaming recommendation.
    \item We conduct extensive offline and online experiments to validate our Moment\&Cross, which has now been deployed on the live-streaming service at Kuaishou, serving 400 Million users.
\end{itemize}

\begin{figure*}[t!]
\begin{center}
\includegraphics[width=18cm,height=9cm]{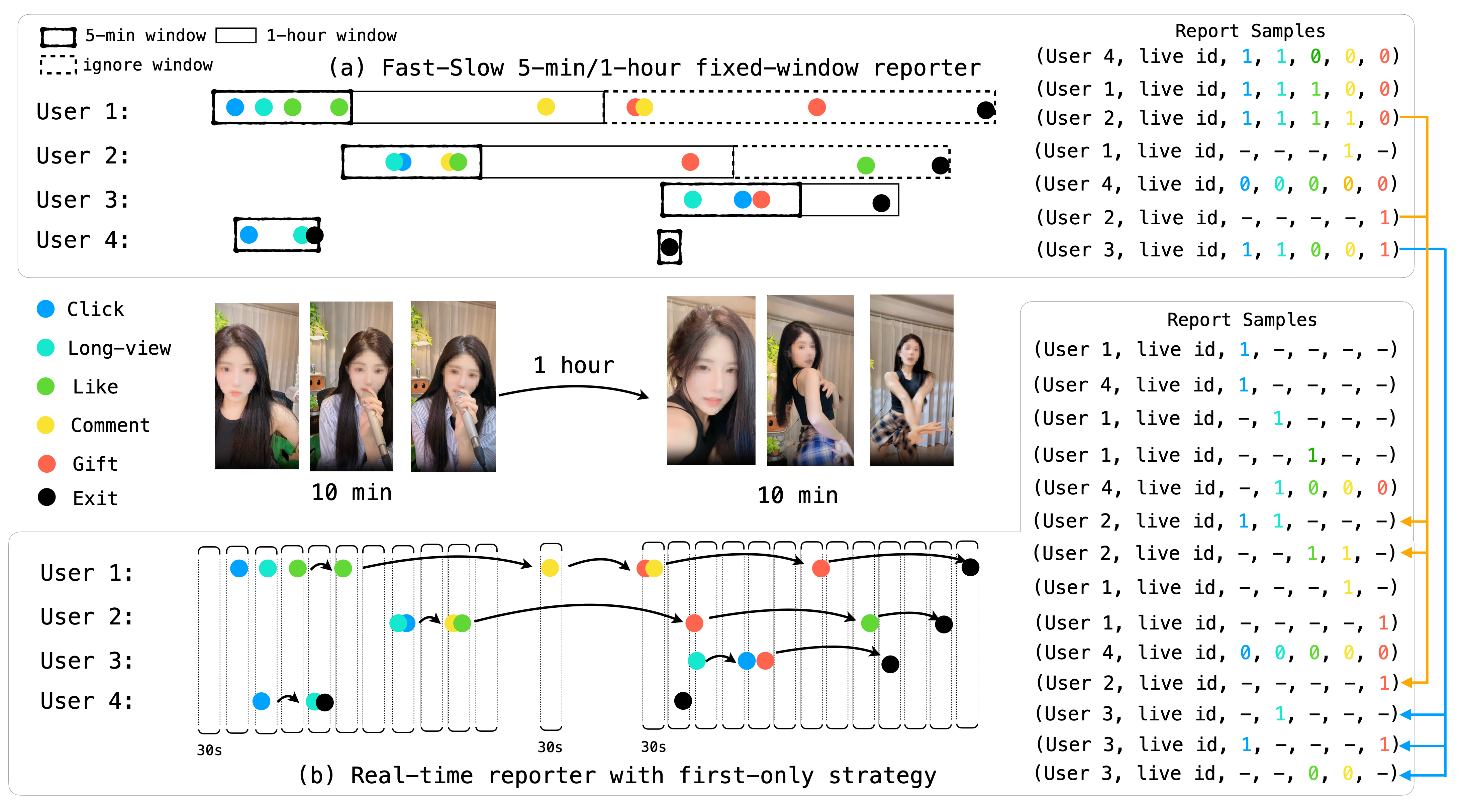}
\caption{The report samples difference of produced training samples between fast-slow 5-min\&1-hour data-steaming and real-time 30s data-steaming. We only show the simplest sample format (user, live-streaming, click, long-view, like, comment, gift). Specifically, for the fast-slow data-streaming, the fast flow reports 5-min window observed all user behaviors, the slow flow reports 5-min missing but 1-hour observed positive user behaviors. In our real-time data-streaming, we report users' first positive behaviors immediately every 30 seconds and report all negative behavior when user exit a live-streaming. According to the report samples' indicative relationship, our real-time data-streaming could produce training samples as soon as possible, to encourage model capturing the CTR increasing trends live-streaming.} 
\label{report_diff}
\end{center}
\end{figure*}

\section{Moment\&Cross at Kuaishou}
The industry CTR prediction model~\cite{widedeep} training process consists of two essential components: (1) a data-streaming engine to organize training samples features and labels,  (2) a multi-task learning based model~\cite{esmm} to fit real training samples' interactions (e.g., click, like, long-view and others).
In this Section, we first review the general preliminaries of our previous 5-min Fast-Slow live-streaming data-streaming engine building and CTR prediction model learning paradigm.
We then show our novel real-time 30s data-streaming engine and its first-only mask learning strategy.
Finally, we give our cross-domain techniques to capture the long-term and short-term short-video interaction patterns of users.

\subsection{Preliminary: Fast-Slow 5-Min\&1-Hour Data-Streaming}
Data-streaming engines as the fundamental component of industry RecSys, a naive solution is to collect the user action logs to report after the item has been fully consumed, e.g., watch and swipe to next short-video, finish a listening song.
Actually, the above solution is `real-time' enough for short-video service, since user exit a short-video average within 1 minute, thus all interaction feedback will be collected in a short time.
Nevertheless, in live-streaming service, users may watch for a very long time (e.g., 30min or 80min), if we still collect all actions when the user exits live-streaming, this will result in the model training not being `real-time' enough.
Consequently, to our knowledge, many live-streaming data-streaming engines follow the \textbf{fixed-window} style to report and assemble training samples features and labels, e.g., 5-min report window.
As one of the characteristics of live-streaming service, different behaviors occur with large different time distributions, and some valuable interactions are hard to observe within a small fixed-window, such as users gifting the author after watching for half an hour.
To this end, we further extend the fixed-window style to the fast-slow windows for our live-streaming service to achieve a balance in Figure~\ref{report_diff}, i.e., \textbf{a fast-window reports all interactions for fast training, and a slow-window reports positive samples that have not been observed in the fast-window}.

Through statistics, we found that most users watch for less than 1 hour, thus we divided the user's watching experience into three periods to monitor user behaviors to supervise our model training:
\begin{itemize}[leftmargin=*,align=left]
    \item \textit{5-minutes}: a small window to collect \textbf{observed all positive and negative behaviors as labels}, to obtain a fast data-streaming.
    \item \textit{1-hour}: a large window only report the \textbf{missing positive labels} to correct the 5-minutes error as a slow data-streaming.
    \item \textit{Ignored window}: this window \textbf{no longer report} any labels.
\end{itemize}

\subsection{Preliminary: CTR Model Training with Positive-Unlabeled Learning}

In a broad sense, the CTR prediction~\cite{dien} model is at the last point of RecSys~\cite{fm}, to rank the most related dozens of items for each user, also called fullrank model.
Indeed, the fullrank~\cite{fuxictr} model is not only to predict the probability that a user would \textit{click} the item candidate (i.e. \textbf{CTR}), and also predict the \textit{long-view} probability (i.e. \textbf{LVTR}), \textit{like} probability (i.e. \textbf{LTR}), \textit{comment} probability (i.e. \textbf{CMTR}) and others \textbf{XTRs} at same time.
According to these predicted probabilities, we can write some complex weighted calculations on these probabilities to control the final score to rank items.

Generally, the fullrank model learning process is formulated as a multi-task~\cite{MTLSurvey} binary classifications paradigm, where the goal is to learn a prediction function $f_\theta(\cdot)$ given the data-streaming training sample.
For each sample consists of the user-item ID,  features raw features $V$ and several real labels to denote the action happens or not (i.e., $y^{ctr}\in \{0,1\}$, $y^{lvtr}\in \{0,1\}$, $y^{ltr}\in \{0,1\}$, $y^{cmtr}\in \{0,1\}$ and others).
Specifically, the raw features $V$ mainly divide four types: user/item IDs, statistics/categories, historical interaction sequences and pre-trained LLM multi-modal~\cite{gpt,flamingo} embeddings, and these features are projected into a low-dimensional embedding as $\mathbf{V} = [\mathbf{v}_1, \mathbf{v}_2, \dots, \mathbf{v}_n]$, $n$ indicates the number of features.
In our live-streaming model, we hand-craft about $n > 400$ raw features to represent user, item and context status.
According to the input sample features and label, the model learning process is formed as:
\begin{equation}
\begin{split}
\hat{y}^{ctr}, \hat{y}^{lvtr}, \hat{y}^{ltr}, \dots = f_\theta([\mathbf{v}_1, \mathbf{v}_2, \dots, \mathbf{v}_n])
\end{split}
\label{ctrprediction}
\end{equation}
where the $\hat{y}^{ctr}, \hat{y}^{lvtr}, \hat{y}^{ltr}, \dots$ are predicted probabilities, while the $f_\theta(\cdot)$ is a multi-task module can be implemented by MMoE~\cite{mmoe}, PLE~\cite{ple}.
Next, we utilize the users' real behavior to supervise our model to optimize parameters.
For the 5-min fast-flow samples, since they report all observed positive and negative labels, thus we train our model by minimizing the standard negative log-likelihood:
\begin{equation}
\begin{split}
\mathcal{L}_{fast} = - \sum_{{ctr, \dots}}^{xtr} \big(y^{xtr}\log{(\hat{y}^{xtr})} + (1-y^{xtr})\log{(1-\hat{y}^{xtr}})\big)
\end{split}
\label{logloss}
\end{equation}
For the 1-hour slow-flow samples, which only report the missing positive labels while masking other consistent positive labels, we employ the positive-unlabeled loss~\cite{puloss,dfm,huangfu2022multi} to \textbf{retract the past "fake negative" error gradient} as:
\begin{equation}
\begin{split}
\mathcal{L}_{slow} = - \sum_{missing}^{xtr} \big(\log{(\hat{y}^{xtr})} - \log{(1-\hat{y}^{xtr}})\big)
\end{split}
\label{puloss}
\end{equation}
where $missing$ represents the positive labels observed only in the 1-hour window.
Through the collaboration of these two losses $\mathcal{L}_{fast}$ and $\mathcal{L}_{slow}$, our model achieves a balance training between effective and efficiency for live-streaming service.
After the model convergence, we can push it as an online fullrank model to respond to real user requests and select the highest score items as follows:
\begin{equation}
\small
\begin{split}
 \texttt{Ranking\_Score} = (1 + \hat{y}^{ctr})^\alpha * (1 + \hat{y}^{lvtr})^\beta * (1 + \hat{y}^{ltr})^\gamma * \dots
\end{split}
\label{ranking}
\end{equation}
where $\alpha$, $\beta$, $\gamma$ are hyper-parameter to assemble all predicted probabilities as one ranking score to sort dozens of item candidates, here we only show a naive case of the weighted \texttt{Ranking\_Score}.

\begin{figure*}[t]
\begin{center}
\includegraphics[width=18cm,height=6.5cm]{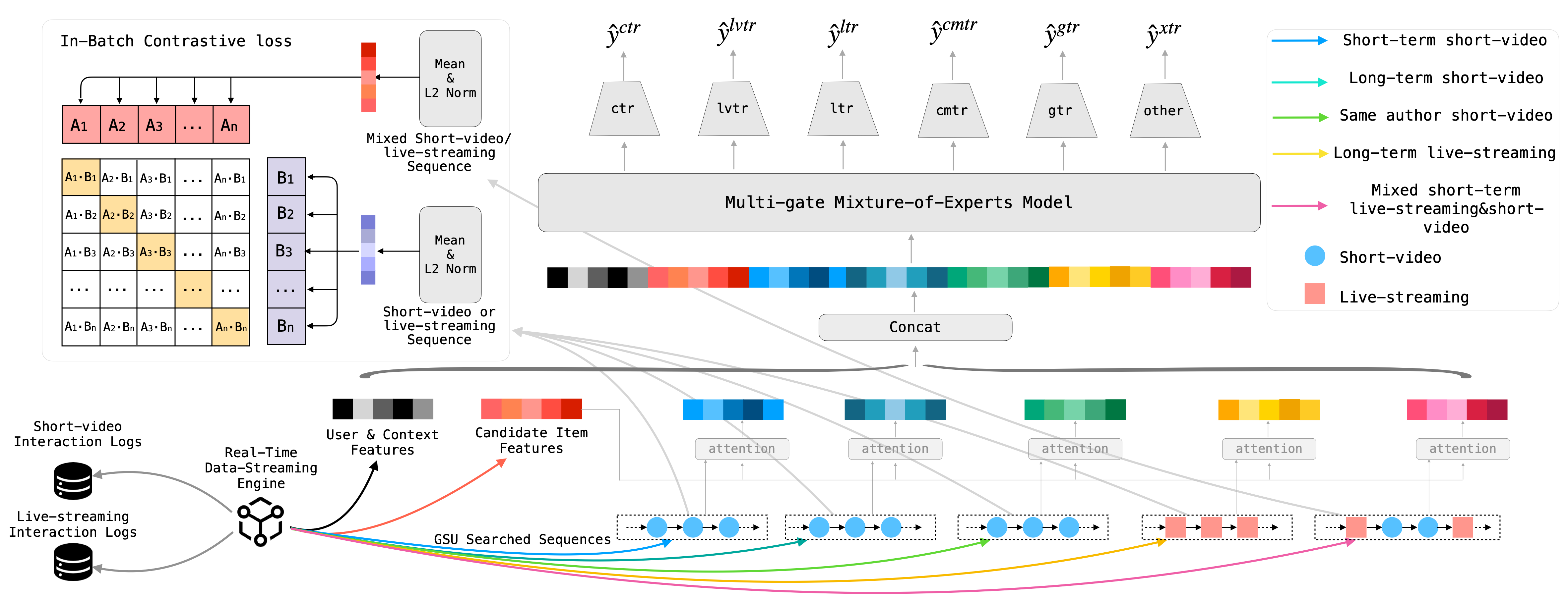}
\caption{We introduce 5 short-video and live-streaming searched sequences to support our model: (1) we first conduct the contrastive mechanism to align them embedding space, (2) we then utilize the target attention mechanism to extract users' interests. (3) we finally concatenate the cross-domain short-video signal to predict each behavior probability.}
\label{modelstructure}
\end{center}
\end{figure*}

\subsection{Moment: Real-time 30s Data-Streaming with First-Only Label-Mask Learning}
As our former version, the Fast-Slow 5-min\&1-hour data-streaming and positive-unlabeled learning framework have been iterating for several years, which is a stable, reliable and proven learning framework designed for live-streaming.
Although effective, it still has some drawbacks, for example: a smaller fixed window will inevitably miss some valuable positive labels, e.g., gift. 
We give the label consistency comparison between fast 5-min and sLow 1-hour data-streaming in Table~\ref{5min1hour}, which describes the proportion of positive samples for major behavior (e.g., click, like, etc.).
We can find that sparser behaviors have lower coverage of label consistency, especially the gift and gift price.

\begin{table}[t]
\footnotesize
\caption{The 5-Min\&1-Hour data streaming label consistency.}
\setlength{\tabcolsep}{5pt}{
\begin{tabular}{ccccccc}
\toprule
&Click &Long-View &Like &Comment & Gift & Gift Price\\ 
\midrule
5-Min/1-Hour &96\% &100\% &79\% &78\% &68\% &34\% \\
\bottomrule
\end{tabular}
}
\label{5min1hour}
\end{table}

Furthermore, the fast 5-min window is still not real-time enough to support our model to capture the CTR increasing trend to solve the serious challenge: \textit{how do we recommend the live-streaming at right moment for users?} 
%
%
%
%
%
%
%
%
To this end, we have upgraded our training framework from Fast-Slow report manner to real-time 30s report manner, to enable our model to perceive all behaviors occurring as soon as possible.
Ideally, if a live-streaming at the `high-light moment', \textbf{there will be a large number of positive gradients to optimize our model parameters in a short period, thus our model can know which live-streaming is in the CTR increasing status and then improve such live-streaming online CTR prediction scores to make it recommended for more users to watch}.
However, the extremely small 30s window may bring some mismatched risks with our former data-streaming as:
\begin{itemize}[leftmargin=*,align=left]
    \item Fake negative: compared with 5-min fixed-window, if we also report all positive\&negative user behaviors after watching 30s, will be introducing to lot of `fake negative' labels since some behaviors are delayed rather than no-occurrence.
    \item Frequent report: compared the fast-slow 5-min\&1-hour data-streaming are only report only once positive label for each behavior. our 30s real-time flow may be reported multiple times (e.g., a user could comment several times).
    \item Interaction beyond: the 30s real-time data-streaming may divide positive labels into several training samples in chronological order. In this way, the early positive behavior updated gradients may help predict the subsequent positive behaviors, for example: the early click positive sample and long-view positive sample updated the model parameters two times, which may overestimate the predicted probability of latter like or comment behaviors.
\end{itemize}
To overcome the fake negative problem, inspired by the \textbf{"mask label"} idea of our former slow 1-hour flow, we introduce a reporting mechanism is that: the positive labels are reported immediately, while the rest negative labels are reported when user exits live-streaming.
As a result, we found that although the report window is much smaller (i.e., 5-min $\rightarrow$ 30s), the amount of data samples did not increase significantly (e.g, about 2$\times$ than Fast-Slow data-streaming), since the additional report samples mainly rely on the sparse behaviors interaction numbers, e.g., like, comment and gift.
For the frequent report problem, we further introduce a \textbf{first-only mask strategy that we only learn the first positive occurrence for each behavior}, to align the learning rule with our former data-streaming.
Therefore, our \textbf{Moment} first-only label-mask learning can be formed as:
\begin{equation}
\begin{split}
\mathcal{L}_{moment} = -\!\!\!\!\!\!\!\!\sum_{{first,exit}}^{xtr} \!\!\!\!\!\!\big(y^{xtr}\log{(\hat{y}^{xtr})} + (1-y^{xtr})\log{(1-\hat{y}^{xtr}})\big)
\end{split}
\label{30sloss}
\end{equation}
where the $first$ indicates the first positive labels of each behavior where others are masked, and the $exit$ denotes the rest of behaviors' positive or negative labels when user exit a live-streaming.
Under the label mask setting, we can utilize Eq.(\ref{30sloss}) to replace Eq.(\ref{logloss}) and Eq.(\ref{puloss}) to support our model training and without the \textbf{long-term feedback delay} issue.
For the interaction beyond risk, particularly, we did not observe this phenomenon, we assume the reason is that the model parameters optimizing is to \textbf{fit all users' data distribution and is hard to overfit to a specific user live-streaming pattern}.
The difference report between fast-slow and 30s real-time data-streaming is shown in Figure~\ref{report_diff}.

\subsection{Cross: Short-Video Interest Transfer}
On the slide page of our model deployment, the exposure content is about 90\% are short-videos with 10\% live-streaming, due to the uneven distribution of our traffic, we must consider the challenge: \textit{how do we utilize users short-video behaviors to make live-streaming recommendation better?}
As shown in Figure~\ref{recarchitecture}, the different business models are only allowed to consume their own training data-streaming, thus our live-streaming model can only be supervised by the users’ live-streaming behaviors.
However, fortunately, we have constructed historical storage services to save users’ interaction logs, and the data-streaming engine could send requests to obtain users’ interaction history from other businesses to assemble them as a part of input features.
In fact, in the `Interaction log', we could retrospect the latest 10,000 watching item ID and access some side-information, e.g., time gap, content multi-modal tag, labels, etc.
In order to model such a long sequence, a wide-used solution is the two cascading search-then-extract~\cite{sim,twin,twinv2} idea: (1) introduce General Search Units (GSUs) to search user history and then filter to obtain a top hundreds related item sequence; (2) devise Exact Search Units (ESUs) to aggregate sequence information to compress users' interests, e.g, sequence pooling, target-item-attention.
In our implementation, we introduce several GSUs to search different interactions from multi-aspects to find the related short-video with target live-streaming candidate, including:
\begin{itemize}[leftmargin=*,align=left]
\item \textit{Latest short-term short-video GSU} aims to search user's hundreds of latest short-video interactions $V^{short}$, which can reflect a precise user's short-term interest point.
\item Dot-product search long-term short-video GSU search the short-video with the highest embedding similarity to the live-streaming candidate, denoted $V^{long}$, which could reflect whether users like this type of live-streaming.
\item Author ID hard search short-video GSU aims to search the short-video history that with the same author ID,  denoted $V^{aidhard}$, which can reflect precise user's interests of this author.
\item Dot-product search long-term live-streaming GSU to obtain $V^{livelong}$, which could reflect whether users like this type of live-streaming according to similar short-video behaviors.
\item Long-view behavior latest mixed live-streaming\&short-video GSU, a positive long-view action hard search to obtain interested live-streaming and short-videos as mixed sequences $V^{mixed}$.
\end{itemize}
For notation brevity, we use the $\mathbf{V}^{short}\in \mathbb{R}^{L*D}$, $\mathbf{V}^{long}\in \mathbb{R}^{L*D}$, $\mathbf{V}^{aidhard}\in \mathbb{R}^{L*D}$, $\mathbf{V}^{livelong}\in \mathbb{R}^{L*D}$, $\mathbf{V}^{mixed}\in \mathbb{R}^{L*D}$ to represent different GSU sequences embeddings, where $L$ is the sequence length.
After obtaining sequence embeddings, we first conduct contrastive objectives to align their embedding spaces:
\begin{equation}
\footnotesize
\begin{split}
\mathcal{L}_{short}^{cl} &= \texttt{Contrastive}\big(\texttt{L2}(\texttt{Mean}(\mathbf{V}^{mixed})), \texttt{L2}(\texttt{Mean}(\mathbf{V}^{short}))\big)\\
\mathcal{L}_{long}^{cl} &= \texttt{Contrastive}\big(\texttt{L2}(\texttt{Mean}(\mathbf{V}^{mixed})), \texttt{L2}(\texttt{Mean}(\mathbf{V}^{long}))\big)\\
\mathcal{L}_{aidhard}^{cl} &= \texttt{Contrastive}\big(\texttt{L2}(\texttt{Mean}(\mathbf{V}^{mixed})), \texttt{L2}(\texttt{Mean}(\mathbf{V}^{aidhard}))\big)\\
\mathcal{L}_{livelong}^{cl} &= \texttt{Contrastive}\big(\texttt{L2}(\texttt{Mean}(\mathbf{V}^{mixed})), \texttt{L2}(\texttt{Mean}(\mathbf{V}^{livelong}))\big)\\
\end{split}
\label{aggregate}
\end{equation}
where the $\texttt{Mean}(\cdot): \mathbb{R}^{L*D}\to \mathbb{R}^{D}$ is a simple pooling function to compress sequence representation, the $\texttt{L2}(\cdot)$ denote L2 normalization function, the $\texttt{Contrastive}(\cdot, \cdot)$ is a \textbf{in-batch sampling} function to gather negative samples to contrastive with.
Inspired by C$^2$DSR~\ref{}, we figure that the mixed live-streaming\&short sequences can be the cornerstone to align with others since it has some similarity with other sequences, but is not completely identical.
Afterward, we conduct the target-item-attention as ESU module to achieve a fine-grained interests extraction according to target live-streaming candidate embedding $\mathbf{V}^{live}$:
\begin{equation}
\footnotesize
\begin{split}
\mathbf{V}^{\cdot}_{ESU} = \texttt{target-item-attention}(&\mathbf{V}^{live}\mathbf{W}^q, \mathbf{V}^{\cdot}\mathbf{W}^k, \mathbf{V}^{\cdot}\mathbf{W}^v),
\end{split}
\label{aggregate}
\end{equation}
where the $\mathbf{V}^{live}$ denotes all live-streaming side features of a training sample (e.g. item tags, Live ID, Author ID, etc.).
After obtaining the enhanced cross-domain short-video interests, we concatenate them to estimate each interaction probability, as shown in Figure~\ref{modelstructure}.

\begin{table*}[t!]
\centering
\caption{Offline Moment\&Cross results(\%) in term AUC of GAUC on live-streaming services at Kuaishou.}
\resizebox{\linewidth}{!}{
\begin{tabular}{ccccccccccccc}
\toprule
\multirow{2}{*}{\makecell{Model\\Variants}} & \multicolumn{2}{c}{Click}  & \multicolumn{2}{c}{Effective-view} & \multicolumn{2}{c}{Long-view} & \multicolumn{2}{c}{Like} & \multicolumn{2}{c}{Comment} & \multicolumn{2}{c}{Gift}
\\ \cmidrule(r){2-3} \cmidrule(r){4-5} \cmidrule(r){6-7} \cmidrule(r){8-9} \cmidrule(r){10-11} \cmidrule(r){12-13}& AUC & GAUC & AUC & GAUC & AUC & GAUC & AUC & GAUC & AUC & GAUC & AUC & GAUC\\
\midrule
PLE (Moment\&Cross) & 82.75 & 65.57 & 78.99 & 64.96 & 85.08 & 75.71 & 91.48 & 74.67 & 92.78 & 74.74 & 96.32 & 74.71        \\
\midrule
CGC (Moment\&Cross) & -0.18 & -0.48 & -0.13 & -0.21 & -0.23 & -0.32 & -0.13 & -0.53 & -0.06 & -0.33 & -0.01 & -0.52       \\
\midrule
Cross w/o $\mathbf{V}^{short}$ & -0.90 & -1.59 & -0.25 & -0.28 & -0.51 & -2.11 & -1.23 & -2.59 & -0.92 & -2.97 & -0.86 & -4.70       \\
Cross w/o $\mathbf{V}^{long}$ & -0.14 & -0.36 & -0.19 & -0.39 & -0.04 & -0.22 & -0.13 & -0.33 & -0.19 & -0.57 & -0.05 & -0.24       \\
Cross w/o $\mathbf{V}^{aidhard}$ & -0.17 & -0.40 & -0.10 & -0.19 & -0.20 & -0.31 & -0.08 & -0.17 & -0.13 & -0.26 & -0.07 & -0.50       \\
Cross w/o $\mathbf{V}^{livelong}$ & -0.11 & -0.25 & -0.08 & -0.11 & -0.10 & -0.11 & -0.25 & -0.55 & -0.05 & -0.16 & -0.02 & -0.15       \\
Cross w/o $\mathbf{V}^{mixed}$ \& $\mathcal{L}^{cl}$ & -0.14 & -0.40 & -0.01 & -0.20 & -0.04 & -0.54 & -0.24 & -0.42 & +0.24 & +0.12 & +0.02 & -0.79      \\
\bottomrule
\end{tabular}
}
\label{mainoffline}
\end{table*}

\section{Experiments}
In this section, we conduct detailed offline experiments and online A/B test on our Kuaishou live-streaming services, to evaluate our proposed method Moment\&Cross.

\subsection{Base Models and Evaluation Metrics}
As shown in Figure~\ref{modelstructure}, in industry ranking model, the multi-gate mixture-of-experts model plays a vital role in estimating various interactions probabilities, and it has several choices, e.g., MMoE~\cite{mmoe}, CGC~\cite{ple}, PLE~\cite{ple}, AdaTT~\cite{adatt} and so on.
In this paper, we select representative multi-task learning methods, \textbf{CGC} and \textbf{PLE}, to verify our solution effectiveness.
We conduct extensive experiments to test our Moment\&Cross ability at live-streaming services and we utilize two classic offline metrics to evaluate our ranking quality, AUC and GAUC~\cite{din} (user grouped AUC).
After our model convergence, we further push it to our online A/B test platform to process real user requests at two applications, Kuaishou and Kuaishou Lite, and we report some major metrics to show our Moment\&Cross improvements, e.g., Watch Time.

\begin{table*}[t!]
\centering
\setlength{\tabcolsep}{8pt}{
\caption{Online Moment\&Cross A/B testing performance of live-streaming services at Kuaishou.}
\begin{tabular}{lcccccccc}
\toprule
\multirow{2}{*}{Applications} &\multirow{2}{*}{Modifications} &\multirow{2}{*}{Groups}                    & \multicolumn{3}{c}{Core Metrics}                    & \multicolumn{3}{c}{Interaction Metrics}                                                                    \\
\cmidrule(r){4-6}  \cmidrule(r){7-9}
                                     &&& Click   & Watch Time                   & Gift Count                     & Like                   & Comment                  & Follow                      \\
\midrule
\multirow{5}{*}{Kuaishou} &Moment & Total  &\textbf{+1.63\%} & \textbf{+4.13\%} & -0.55\% & - & - & \textbf{+3.70\%} \\
\cmidrule(r){2-9}
\multirow{5}{*}{}  &\multirow{4}{*}{Corss} & Total   &  \textbf{+2.21\%} & \textbf{+2.27\%} & \textbf{+6.91\%} & \textbf{+2.84\%} & \textbf{+2.23\%} & \textbf{+4.21\%}\\
\cmidrule(r){3-9}
\multirow{5}{*}{} & \multirow{4}{*}{} & Low-Gift  &  \textbf{+6.75\%} & \textbf{+6.56\%} & \textbf{+8.50\%} & \textbf{+0.53\%} & \textbf{+7.75\%} & \textbf{+12.84\%}\\

\multirow{5}{*}{}  &\multirow{4}{*}{} & Middle-Gift   &  \textbf{+4.27\%} & \textbf{+3.38\%} & \textbf{+9.21\%} & \textbf{+3.77\%} & \textbf{+6.47\%} & \textbf{+7.08\%}\\

\multirow{5}{*}{}  &\multirow{4}{*}{} & High-Gift    &  \textbf{+0.11\%} & \textbf{+1.16\%} & \textbf{+4.15\%} & \textbf{+3.36\%} & \textbf{+0.15\%} & \textbf{+0.26\%}\\

\midrule
\multirow{5}{*}{\makecell{Kuaishou\\Lite}} &Moment & Total  &\textbf{+0.64\%} & \textbf{+1.85\%} & -1.22\% & - & - & \textbf{+2.79\%} \\
\cmidrule(r){2-9}
\multirow{5}{*}{} &\multirow{4}{*}{Cross} & Total   &  \textbf{+2.72\%} & \textbf{+2.48\%} & \textbf{+8.91\%} & \textbf{+1.05\%} & \textbf{+4.58\%} & \textbf{+4.37\%}\\
\cmidrule(r){3-9}
\multirow{5}{*}{} &\multirow{4}{*}{} & Low-Gift  &  \textbf{+2.94\%} & \textbf{+5.69\%} & \textbf{+24.75\%} & \textbf{+5.75\%} & \textbf{+3.94\%} & \textbf{+7.07\%}\\
\multirow{5}{*}{} &\multirow{4}{*}{} & Middle-Gift    &  \textbf{+5.35\%} & \textbf{+3.86\%} & \textbf{+15.08\%} & \textbf{+2.41\%} & \textbf{+3.11\%} & \textbf{+8.76\%}\\

\multirow{5}{*}{} &\multirow{4}{*}{} & High-Gift   &   \textbf{+0.52\%} & \textbf{+1.93\%} & \textbf{+4.42\%} & \textbf{+0.21\%} & \textbf{+4.18\%} & \textbf{+1.37\%}\\
\bottomrule
\label{mainonline}
\end{tabular}
}
\vspace{-0.1cm}
\end{table*}

\subsection{Overall Performance}
Table~\ref{mainoffline} shows the performances of our Moment and Cross, respectively.
Specifically, our online service needs to tackle billions of user requests per day, and the improvement of 0.10\% in offline evaluation of AUC and GAUC is significant enough to bring online gains.
From the results, we have the following observations:

\begin{itemize}[leftmargin=*,align=left]
    \item 
    (1) To investigate our real-time data-streaming effectiveness, we implement two multi-task variants: PLE(Moment\&Cross) and CGC(Moment\&Cross), where PLE is a double-layer stacked version of CGC, which is also the deployed model to support online request traffic.
    According Table~\ref{mainoffline}, we can observe the PLE variant shows the expected confidence improvement compared to the CGC variant, which reveals our new real-time data-streaming could seamlessly support other models with the first-only label-mask learning strategy.
    \item To validate the Cross domain short-video interest effectiveness, we conduct ablation studies to test all GSU sequences effectiveness one by one, e.g., Cross without short-term short-video sequence.
    From Table~\ref{mainoffline}, we have the following observations: 
    (1) All of our Cross variants show significant performance degeneration, which indicates users' history short-video or live-streaming sequences could enhance our model to capture users interests more accurately. 
    (2) Compared with the live-streaming based sequence, the short-video based sequences are more powerful in providing predictive information to empower live-streaming ranking model, i.e., the short-term short-video sequence $\mathbf{V}^{short}$ could provide 0.9\% improvement. 
    The reason might be that users watching experience about 90\% of the watching content are short-videos, leading their interest points will be better reflected in short-video historical sequence.
    which reveals that transferring the user cross-domain interest from rich short-video domain is a powerful technique to support our live-streaming service.
\end{itemize}

\subsection{Online A/B Test}
To quantify the contribution of Moment\&Cross could brings to our live-streaming services, we push the corresponding modification to our online A/B test system to serve as a ranking model at two applications, the Kuaishou and Kuaishou Lite.
We evaluate model performance based on the core metrics and interaction metrics, e.g., Watch-Time, Gift Price, Click, etc.
Table~\ref{mainonline} reports our online results of Moment and Cross individually, here we further shows the fine-grained improvement value of three user groups to test our short-video interests transferring: Low/Middle/High-Gift users and total users.
Specifically, since a long time has passed of our Moment modification, we unfortunately lost the likes and comments interaction metric results, but our core metrics were retained.
According Table~\ref{mainonline}, we can find real-time Moment data-streaming trained model achieves a large improvement at Click +1.63\%/+0.64\% and Watch Time +4.13\%/+1.85\%, which indicates that accelerating the model real-time training efficiency is crucial for live-streaming recommendation.
Regarding the slightly negative -0.55\%/-1.22\% of Gift Count, this is because gift is an \textbf{unstable metric} and within a reasonable range in our system.
Our Cross achieves a large improvement of $2.27\%$/$2.48\%$ at watch-time and $6.91\%$/$8.91\%$ at gift price, the results demonstrate that our Cross could contribute to our system significantly.
Further, the Low-Gift user group shows the biggest growth than others, which indicates the rich short-video signal is helpful for our system to alleviate data-sparse problem.

\begin{figure}[t]
\begin{center}
\includegraphics[width=7cm,height=6cm]{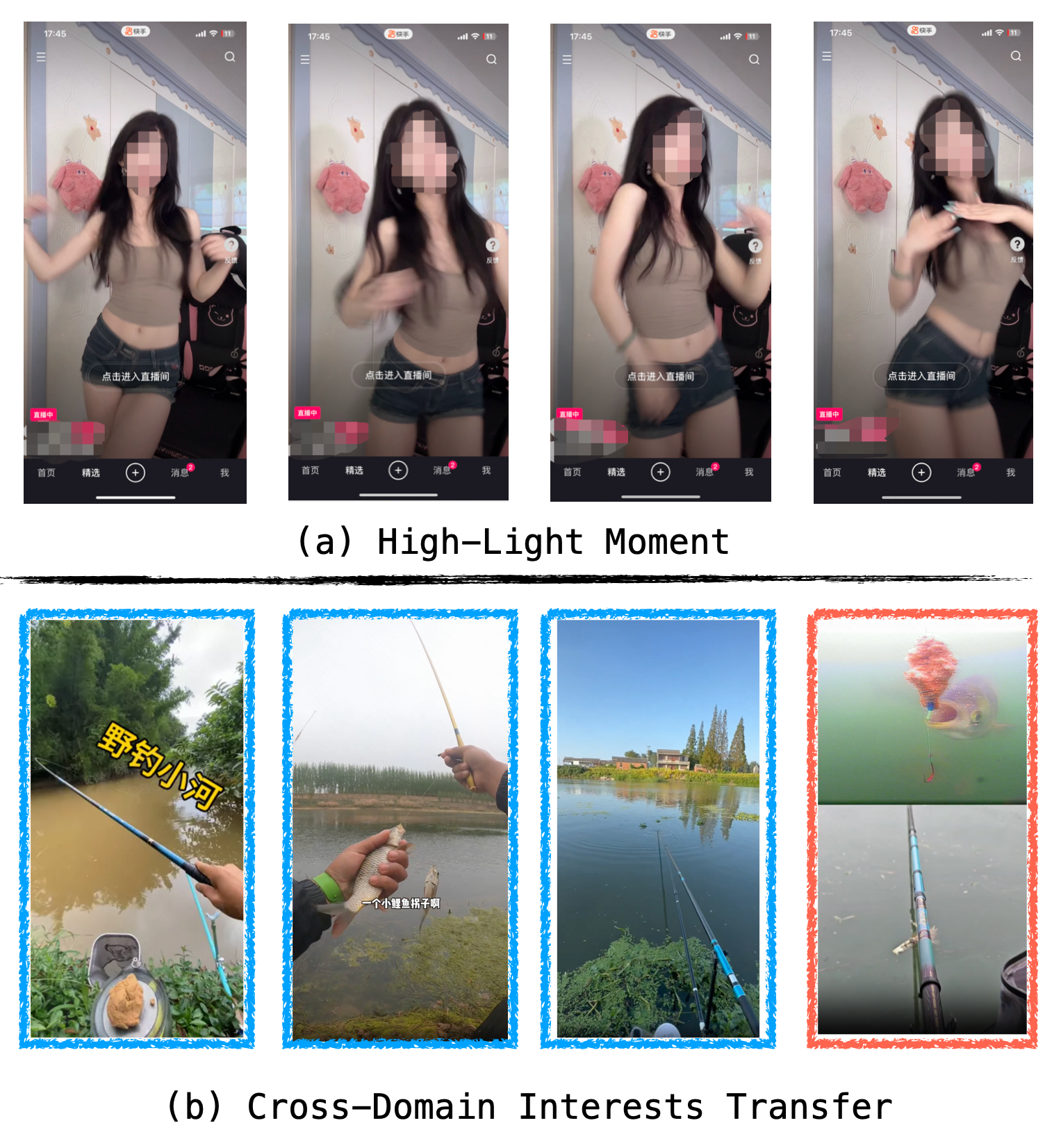}
\caption{(a) Moment could enhance our model to perceive the high-light live-streamings; (b) Short-video cross-domain interests could help our system find related live-streamings.}
\label{livecase}
\end{center}
\end{figure}

\subsection{Case Study}
This Section discovers the certain experience influence of Moment\&Cross, we give three cases to demonstrate its effectiveness:
\begin{itemize}[leftmargin=*,align=left]
\item As shown in Figure~\ref{livecase}(a), we find out that the Slide page could recommend more `high-light moment' of talent-show authors, which indicates the 30s real-time data-streaming with mask learning paradigm could capture the CTR increasing trends to detect `high-light moment' accurately.
And such phenomenon will not only bring a better experience to user but also allow the efforts of our platform's authors to be seen by more users, building a better environment for live-streaming.
\item For the cross-domain interests transferring, as shown in Figure~\ref{livecase}(b), we observe that our system could feed more related live-streamings for users, e.g., a user long-viewed some fishing videos and then recommend outdoor fishing live-streamings. To be specific, fishing live-streaming is a minority category, without the help of short-video signals, it is hard for our system to successfully recommend this live-streaming.
\item Moreover, since Kuaishou is a mainly short-video platform, many users do not have the habit of watching live-streaming.
We are also curious whether our model can encourage more users to watch live-streaming with the help of short video signals.
Here we divide our users into four types (e.g., low/middle/high/full-activate user groups) to show our model's predictive ability.
From Table~\ref{ablation}, we can find the low-activate group shows the most significant improvements than other groups, which indicates our cross-domain interest transfer could effectively discover potential user group for live-streaming service.
\end{itemize}

\section{Related Works}
In recent years, live-streaming has become a fashionable phenomenon, with a large number of professional authors relying on live-streaming media to interact with the audience.
Different from other recommendation scenarios that connect users with some items, the live-streaming aims to link users with their interested author, thus the pioneer work LiveRec~\cite{liverec} considering the user-author repeated consumption relationship with self-attention mechanism.
Further, users could watch a live-streaming for a long time, the \cite{chen2022weighing} devise a loss re-weighted strategy, which adjusts loss by differing amounts of minutes watched.
To consider the live-streaming multi-modal information effects, the MTA~\cite{xi2023multimodal} and ContentCTR~\cite{contentctr} further introduce some multi-modal components to fuse textual, image frame information, and the Sliver~\cite{sliver} introduces a re-recommendation mechanism to capture dynamic live-streaming change.
Besides, the recent aims to capture the user-author and author-author relation, MMBee~\cite{deng2024mmbee} provides a graph representation learning and meta-path based behavior expansion strategy to enrich user and item multi-hop neighboring information.
In addition, the recent progress also points out the cross-domain~\cite{cdrib, disencdr, unicdr} signal ability for live-streaming, the DIAGAE~\cite{zheng2023dual} utilizes live-streaming domain user representations to align with other rich services learned user representations.
The eLiveRec~\cite{eliverec} aims to improve e-commerce live streaming recommendation, which devises a disentangled encoder to learn user’s live-streaming and product shared intentions and live-streaming specific intentions.
Compared with them, our Moment\&Cross has the distinct motivations that we aim to solve the following two problems:  (1) \textit{how do we recommend the live-streamings at right moment for users?} (2) \textit{how do we utilize users rich short-video behaviors to make live-streaming recommendation better?}
Further, the used solutions are totally different: (1) we upgrade our data-streaming and devise a novel first-only mask learning strategy, (2) we introduce search-based framework to exploit rich domain interaction sequences with a contrastive objective.

\begin{table}[t]
\centering
\caption{Short-video interest effects of different user group.}
\setlength{\tabcolsep}{9.5pt}{
\begin{tabular}{ccccc}
\toprule
\multirow{2}{*}{Metrics} & \multicolumn{4}{c}{Live Activate User Groups} \\
 \cmidrule(r){2-5} & Low  & Middle & High & Full \\
\midrule
Click & -1.45\% & -1.35\% & -1.24\% & -0.95\%  \\
Long-view & -1.79\% & -1.74\% & -1.74\% & -1.57\%    \\
Like & -1.88\% & -1.97\% & -1.86\% & -1.25\%   \\
Comment & -1.48\% & -1.67\% & -1.64\%  & -1.14\%    \\
Gift & -1.47\% & -1.67\% & -1.53\% & -0.94\%   \\
\bottomrule
\end{tabular}
}
\label{ablation}
\end{table}

\section{Conclusions}
In this paper, we propose the Moment\&Cross, towards to build the next-generation recommendation model for live-streaming services.
Specifically, we first explain the background of live-streaming at Kuaishou, and then present two questions: (1) \textit{how do we recommend the live-streamings at right moment for users?} (2) \textit{how do we utilize users rich short-video behaviors to make live-streaming recommendation better?}
For the first challenge, we blame the reason that 5-min fixed report window is hard to perceive live-streaming hot trends, and we describe a real-time 30s report mechanism with first-only mask learning paradigm to alleviate the problem.
For the second challenge, we first devise several GSUs to search short-video and live-streaming sequences from user historical logs, and then introduce a contrastive objective to align them with representation space to support our ranking model.
Extensive offline and online experimental results on our industrial real-time 30s data-streaming demonstrate the effectiveness of Moment\&Cross at live-streaming services.
Further, detailed analyses from various perspectives show the effectiveness of Moment and Cross, respectively.

\balance
\bibliographystyle{ACM-Reference-Format}
\bibliography{sample-base-extend.bib}


\begin{thebibliography}{39}


\ifx \showCODEN    \undefined \def \showCODEN     #1{\unskip}     \fi
\ifx \showDOI      \undefined \def \showDOI       #1{#1}\fi
\ifx \showISBNx    \undefined \def \showISBNx     #1{\unskip}     \fi
\ifx \showISBNxiii \undefined \def \showISBNxiii  #1{\unskip}     \fi
\ifx \showISSN     \undefined \def \showISSN      #1{\unskip}     \fi
\ifx \showLCCN     \undefined \def \showLCCN      #1{\unskip}     \fi
\ifx \shownote     \undefined \def \shownote      #1{#1}          \fi
\ifx \showarticletitle \undefined \def \showarticletitle #1{#1}   \fi
\ifx \showURL      \undefined \def \showURL       {\relax}        \fi
\providecommand\bibfield[2]{#2}
\providecommand\bibinfo[2]{#2}
\providecommand\natexlab[1]{#1}
\providecommand\showeprint[2][]{arXiv:#2}

\bibitem[Awadalla et~al\mbox{.}(2023)]%
        {flamingo}
\bibfield{author}{\bibinfo{person}{Anas Awadalla}, \bibinfo{person}{Irena Gao}, \bibinfo{person}{Josh Gardner}, \bibinfo{person}{Jack Hessel}, \bibinfo{person}{Yusuf Hanafy}, \bibinfo{person}{Wanrong Zhu}, \bibinfo{person}{Kalyani Marathe}, \bibinfo{person}{Yonatan Bitton}, \bibinfo{person}{Samir Gadre}, \bibinfo{person}{Shiori Sagawa}, {et~al\mbox{.}}} \bibinfo{year}{2023}\natexlab{}.
\newblock \showarticletitle{Openflamingo: An open-source framework for training large autoregressive vision-language models}.
\newblock \bibinfo{journal}{\emph{Arxiv}} (\bibinfo{year}{2023}).
\newblock


\bibitem[Badanidiyuru et~al\mbox{.}(2021)]%
        {handlingdelay}
\bibfield{author}{\bibinfo{person}{Ashwinkumar Badanidiyuru}, \bibinfo{person}{Andrew Evdokimov}, \bibinfo{person}{Vinodh Krishnan}, \bibinfo{person}{Pan Li}, \bibinfo{person}{Wynn Vonnegut}, {and} \bibinfo{person}{Jayden Wang}.} \bibinfo{year}{2021}\natexlab{}.
\newblock \showarticletitle{Handling many conversions per click in modeling delayed feedback}.
\newblock \bibinfo{journal}{\emph{Arxiv}} (\bibinfo{year}{2021}).
\newblock


\bibitem[Belghazi et~al\mbox{.}(2018)]%
        {mine}
\bibfield{author}{\bibinfo{person}{Mohamed~Ishmael Belghazi}, \bibinfo{person}{Aristide Baratin}, \bibinfo{person}{Sai Rajeshwar}, \bibinfo{person}{Sherjil Ozair}, \bibinfo{person}{Yoshua Bengio}, \bibinfo{person}{Aaron Courville}, {and} \bibinfo{person}{Devon Hjelm}.} \bibinfo{year}{2018}\natexlab{}.
\newblock \showarticletitle{Mutual information neural estimation}. In \bibinfo{booktitle}{\emph{International Conference on Machine Learning (ICML)}}.
\newblock


\bibitem[Brown et~al\mbox{.}(2020)]%
        {gpt}
\bibfield{author}{\bibinfo{person}{Tom Brown}, \bibinfo{person}{Benjamin Mann}, \bibinfo{person}{Nick Ryder}, \bibinfo{person}{Melanie Subbiah}, \bibinfo{person}{Jared~D Kaplan}, \bibinfo{person}{Prafulla Dhariwal}, \bibinfo{person}{Arvind Neelakantan}, \bibinfo{person}{Pranav Shyam}, \bibinfo{person}{Girish Sastry}, \bibinfo{person}{Amanda Askell}, {et~al\mbox{.}}} \bibinfo{year}{2020}\natexlab{}.
\newblock \showarticletitle{Language models are few-shot learners}. In \bibinfo{booktitle}{\emph{Conference on Neural Information Processing Systems (NeurIPS)}}.
\newblock


\bibitem[Cao et~al\mbox{.}(2023)]%
        {unicdr}
\bibfield{author}{\bibinfo{person}{Jiangxia Cao}, \bibinfo{person}{Shaoshuai Li}, \bibinfo{person}{Bowen Yu}, \bibinfo{person}{Xiaobo Guo}, \bibinfo{person}{Tingwen Liu}, {and} \bibinfo{person}{Bin Wang}.} \bibinfo{year}{2023}\natexlab{}.
\newblock \showarticletitle{Towards universal cross-domain recommendation}. In \bibinfo{booktitle}{\emph{ACM International Conference on Web Search and Data Mining (WSDM)}}.
\newblock


\bibitem[Cao et~al\mbox{.}(2022a)]%
        {disencdr}
\bibfield{author}{\bibinfo{person}{Jiangxia Cao}, \bibinfo{person}{Xixun Lin}, \bibinfo{person}{Xin Cong}, \bibinfo{person}{Jing Ya}, \bibinfo{person}{Tingwen Liu}, {and} \bibinfo{person}{Bin Wang}.} \bibinfo{year}{2022}\natexlab{a}.
\newblock \showarticletitle{Disencdr: Learning disentangled representations for cross-domain recommendation}. In \bibinfo{booktitle}{\emph{International ACM SIGIR Conference on Research and Development in Information Retrieval (SIGIR)}}.
\newblock


\bibitem[Cao et~al\mbox{.}(2022b)]%
        {cdrib}
\bibfield{author}{\bibinfo{person}{Jiangxia Cao}, \bibinfo{person}{Jiawei Sheng}, \bibinfo{person}{Xin Cong}, \bibinfo{person}{Tingwen Liu}, {and} \bibinfo{person}{Bin Wang}.} \bibinfo{year}{2022}\natexlab{b}.
\newblock \showarticletitle{Cross-domain recommendation to cold-start users via variational information bottleneck}. In \bibinfo{booktitle}{\emph{IEEE International Conference on Data Engineering (ICDE)}}.
\newblock


\bibitem[Cao et~al\mbox{.}(2022c)]%
        {sdim}
\bibfield{author}{\bibinfo{person}{Yue Cao}, \bibinfo{person}{Xiaojiang Zhou}, \bibinfo{person}{Jiaqi Feng}, \bibinfo{person}{Peihao Huang}, \bibinfo{person}{Yao Xiao}, \bibinfo{person}{Dayao Chen}, {and} \bibinfo{person}{Sheng Chen}.} \bibinfo{year}{2022}\natexlab{c}.
\newblock \showarticletitle{Sampling is all you need on modeling long-term user behaviors for CTR prediction}. In \bibinfo{booktitle}{\emph{ACM International Conference on Information and Knowledge Management (CIKM)}}.
\newblock


\bibitem[Chang et~al\mbox{.}(2023a)]%
        {twin}
\bibfield{author}{\bibinfo{person}{Jianxin Chang}, \bibinfo{person}{Chenbin Zhang}, \bibinfo{person}{Zhiyi Fu}, \bibinfo{person}{Xiaoxue Zang}, \bibinfo{person}{Lin Guan}, \bibinfo{person}{Jing Lu}, \bibinfo{person}{Yiqun Hui}, \bibinfo{person}{Dewei Leng}, \bibinfo{person}{Yanan Niu}, \bibinfo{person}{Yang Song}, {et~al\mbox{.}}} \bibinfo{year}{2023}\natexlab{a}.
\newblock \showarticletitle{TWIN: TWo-stage interest network for lifelong user behavior modeling in CTR prediction at kuaishou}. In \bibinfo{booktitle}{\emph{ACM SIGKDD Conference on Knowledge Discovery and Data Mining (KDD)}}.
\newblock


\bibitem[Chang et~al\mbox{.}(2023b)]%
        {pepnet}
\bibfield{author}{\bibinfo{person}{Jianxin Chang}, \bibinfo{person}{Chenbin Zhang}, \bibinfo{person}{Yiqun Hui}, \bibinfo{person}{Dewei Leng}, \bibinfo{person}{Yanan Niu}, \bibinfo{person}{Yang Song}, {and} \bibinfo{person}{Kun Gai}.} \bibinfo{year}{2023}\natexlab{b}.
\newblock \showarticletitle{Pepnet: Parameter and embedding personalized network for infusing with personalized prior information}. In \bibinfo{booktitle}{\emph{ACM SIGKDD Conference on Knowledge Discovery and Data Mining (KDD)}}.
\newblock


\bibitem[Chapelle(2014)]%
        {dfm}
\bibfield{author}{\bibinfo{person}{Olivier Chapelle}.} \bibinfo{year}{2014}\natexlab{}.
\newblock \showarticletitle{Modeling delayed feedback in display advertising}. In \bibinfo{booktitle}{\emph{ACM SIGKDD Conference on Knowledge Discovery and Data Mining (KDD)}}.
\newblock


\bibitem[Chen et~al\mbox{.}(2022)]%
        {chen2022weighing}
\bibfield{author}{\bibinfo{person}{Edgar Chen}, \bibinfo{person}{Mark Ally}, \bibinfo{person}{Eder Santana}, {and} \bibinfo{person}{Saad Ali}.} \bibinfo{year}{2022}\natexlab{}.
\newblock \showarticletitle{Weighing dynamic availability and consumption for Twitch recommendations}.
\newblock \bibinfo{journal}{\emph{Arxiv}} (\bibinfo{year}{2022}).
\newblock


\bibitem[Chen et~al\mbox{.}(2021)]%
        {eta}
\bibfield{author}{\bibinfo{person}{Qiwei Chen}, \bibinfo{person}{Changhua Pei}, \bibinfo{person}{Shanshan Lv}, \bibinfo{person}{Chao Li}, \bibinfo{person}{Junfeng Ge}, {and} \bibinfo{person}{Wenwu Ou}.} \bibinfo{year}{2021}\natexlab{}.
\newblock \showarticletitle{End-to-end user behavior retrieval in click-through rateprediction model}.
\newblock \bibinfo{journal}{\emph{Arxiv}} (\bibinfo{year}{2021}).
\newblock


\bibitem[Cheng et~al\mbox{.}(2016)]%
        {widedeep}
\bibfield{author}{\bibinfo{person}{Heng-Tze Cheng}, \bibinfo{person}{Levent Koc}, \bibinfo{person}{Jeremiah Harmsen}, \bibinfo{person}{Tal Shaked}, \bibinfo{person}{Tushar Chandra}, \bibinfo{person}{Hrishi Aradhye}, \bibinfo{person}{Glen Anderson}, \bibinfo{person}{Greg Corrado}, \bibinfo{person}{Wei Chai}, \bibinfo{person}{Mustafa Ispir}, {et~al\mbox{.}}} \bibinfo{year}{2016}\natexlab{}.
\newblock \showarticletitle{Wide \& deep learning for recommender systems}. In \bibinfo{booktitle}{\emph{ACM Conference on Recommender Systems (RecSys Workshop)}}.
\newblock


\bibitem[Covington et~al\mbox{.}(2016)]%
        {youtube}
\bibfield{author}{\bibinfo{person}{Paul Covington}, \bibinfo{person}{Jay Adams}, {and} \bibinfo{person}{Emre Sargin}.} \bibinfo{year}{2016}\natexlab{}.
\newblock \showarticletitle{Deep Neural Networks for YouTube Recommendations}. In \bibinfo{booktitle}{\emph{ACM Conference on Recommender Systems (RecSys)}}.
\newblock


\bibitem[Deng et~al\mbox{.}(2023)]%
        {contentctr}
\bibfield{author}{\bibinfo{person}{Jiaxin Deng}, \bibinfo{person}{Dong Shen}, \bibinfo{person}{Shiyao Wang}, \bibinfo{person}{Xiangyu Wu}, \bibinfo{person}{Fan Yang}, \bibinfo{person}{Guorui Zhou}, {and} \bibinfo{person}{Gaofeng Meng}.} \bibinfo{year}{2023}\natexlab{}.
\newblock \showarticletitle{ContentCTR: Frame-level Live Streaming Click-Through Rate Prediction with Multimodal Transformer}.
\newblock \bibinfo{journal}{\emph{Arxiv}} (\bibinfo{year}{2023}).
\newblock


\bibitem[Deng et~al\mbox{.}(2024)]%
        {deng2024mmbee}
\bibfield{author}{\bibinfo{person}{Jiaxin Deng}, \bibinfo{person}{Shiyao Wang}, \bibinfo{person}{Yuchen Wang}, \bibinfo{person}{Jiansong Qi}, \bibinfo{person}{Liqin Zhao}, \bibinfo{person}{Guorui Zhou}, {and} \bibinfo{person}{Gaofeng Meng}.} \bibinfo{year}{2024}\natexlab{}.
\newblock \showarticletitle{MMBee: Live Streaming Gift-Sending Recommendations via Multi-Modal Fusion and Behaviour Expansion}. In \bibinfo{booktitle}{\emph{ACM SIGKDD Conference on Knowledge Discovery and Data Mining (KDD)}}.
\newblock


\bibitem[Huangfu et~al\mbox{.}(2022)]%
        {huangfu2022multi}
\bibfield{author}{\bibinfo{person}{Zhigang Huangfu}, \bibinfo{person}{Gong-Duo Zhang}, \bibinfo{person}{Zhengwei Wu}, \bibinfo{person}{Qintong Wu}, \bibinfo{person}{Zhiqiang Zhang}, \bibinfo{person}{Lihong Gu}, \bibinfo{person}{Jun Zhou}, {and} \bibinfo{person}{Jinjie Gu}.} \bibinfo{year}{2022}\natexlab{}.
\newblock \showarticletitle{A multi-task learning approach for delayed feedback modeling}. In \bibinfo{booktitle}{\emph{International World Wide Web Conference Companion}}.
\newblock


\bibitem[Ktena et~al\mbox{.}(2019)]%
        {puloss}
\bibfield{author}{\bibinfo{person}{Sofia~Ira Ktena}, \bibinfo{person}{Alykhan Tejani}, \bibinfo{person}{Lucas Theis}, \bibinfo{person}{Pranay~Kumar Myana}, \bibinfo{person}{Deepak Dilipkumar}, \bibinfo{person}{Ferenc Husz{\'a}r}, \bibinfo{person}{Steven Yoo}, {and} \bibinfo{person}{Wenzhe Shi}.} \bibinfo{year}{2019}\natexlab{}.
\newblock \showarticletitle{Addressing delayed feedback for continuous training with neural networks in CTR prediction}. In \bibinfo{booktitle}{\emph{ACM Conference on Recommender Systems (RecSys)}}.
\newblock


\bibitem[Li et~al\mbox{.}(2023)]%
        {adatt}
\bibfield{author}{\bibinfo{person}{Danwei Li}, \bibinfo{person}{Zhengyu Zhang}, \bibinfo{person}{Siyang Yuan}, \bibinfo{person}{Mingze Gao}, \bibinfo{person}{Weilin Zhang}, \bibinfo{person}{Chaofei Yang}, \bibinfo{person}{Xi Liu}, {and} \bibinfo{person}{Jiyan Yang}.} \bibinfo{year}{2023}\natexlab{}.
\newblock \showarticletitle{AdaTT: Adaptive Task-to-Task Fusion Network for Multitask Learning in Recommendations}. In \bibinfo{booktitle}{\emph{ACM SIGKDD Conference on Knowledge Discovery and Data Mining (KDD)}}.
\newblock


\bibitem[Liang et~al\mbox{.}(2024)]%
        {sliver}
\bibfield{author}{\bibinfo{person}{Fengqi Liang}, \bibinfo{person}{Baigong Zheng}, \bibinfo{person}{Liqin Zhao}, \bibinfo{person}{Guorui Zhou}, \bibinfo{person}{Qian Wang}, {and} \bibinfo{person}{Yanan Niu}.} \bibinfo{year}{2024}\natexlab{}.
\newblock \showarticletitle{Ensure Timeliness and Accuracy: A Novel Sliding Window Data Stream Paradigm for Live Streaming Recommendation}.
\newblock \bibinfo{journal}{\emph{Arxiv}} (\bibinfo{year}{2024}).
\newblock


\bibitem[Ma et~al\mbox{.}(2018b)]%
        {mmoe}
\bibfield{author}{\bibinfo{person}{Jiaqi Ma}, \bibinfo{person}{Zhe Zhao}, \bibinfo{person}{Xinyang Yi}, \bibinfo{person}{Jilin Chen}, \bibinfo{person}{Lichan Hong}, {and} \bibinfo{person}{Ed~H Chi}.} \bibinfo{year}{2018}\natexlab{b}.
\newblock \showarticletitle{Modeling task relationships in multi-task learning with multi-gate mixture-of-experts}. In \bibinfo{booktitle}{\emph{ACM SIGKDD Conference on Knowledge Discovery and Data Mining (KDD)}}.
\newblock


\bibitem[Ma et~al\mbox{.}(2018a)]%
        {esmm}
\bibfield{author}{\bibinfo{person}{Xiao Ma}, \bibinfo{person}{Liqin Zhao}, \bibinfo{person}{Guan Huang}, \bibinfo{person}{Zhi Wang}, \bibinfo{person}{Zelin Hu}, \bibinfo{person}{Xiaoqiang Zhu}, {and} \bibinfo{person}{Kun Gai}.} \bibinfo{year}{2018}\natexlab{a}.
\newblock \showarticletitle{Entire space multi-task model: An effective approach for estimating post-click conversion rate}. In \bibinfo{booktitle}{\emph{International ACM SIGIR Conference on Research and Development in Information Retrieval (SIGIR)}}.
\newblock


\bibitem[Pi et~al\mbox{.}(2020)]%
        {sim}
\bibfield{author}{\bibinfo{person}{Qi Pi}, \bibinfo{person}{Guorui Zhou}, \bibinfo{person}{Yujing Zhang}, \bibinfo{person}{Zhe Wang}, \bibinfo{person}{Lejian Ren}, \bibinfo{person}{Ying Fan}, \bibinfo{person}{Xiaoqiang Zhu}, {and} \bibinfo{person}{Kun Gai}.} \bibinfo{year}{2020}\natexlab{}.
\newblock \showarticletitle{Search-based User Interest Modeling with Lifelong Sequential Behavior Data for Click-Through Rate Prediction}. In \bibinfo{booktitle}{\emph{ACM International Conference on Information and Knowledge Management (CIKM)}}.
\newblock


\bibitem[Rappaz et~al\mbox{.}(2021)]%
        {liverec}
\bibfield{author}{\bibinfo{person}{J{\'e}r{\'e}mie Rappaz}, \bibinfo{person}{Julian McAuley}, {and} \bibinfo{person}{Karl Aberer}.} \bibinfo{year}{2021}\natexlab{}.
\newblock \showarticletitle{Recommendation on live-streaming platforms: Dynamic availability and repeat consumption}. In \bibinfo{booktitle}{\emph{ACM Conference on Recommender Systems (RecSys)}}.
\newblock


\bibitem[Rendle(2010)]%
        {fm}
\bibfield{author}{\bibinfo{person}{Steffen Rendle}.} \bibinfo{year}{2010}\natexlab{}.
\newblock \showarticletitle{Factorization machines}. In \bibinfo{booktitle}{\emph{IEEE International Conference on Data Mining (ICDM)}}.
\newblock


\bibitem[Si et~al\mbox{.}(2024)]%
        {twinv2}
\bibfield{author}{\bibinfo{person}{Zihua Si}, \bibinfo{person}{Lin Guan}, \bibinfo{person}{ZhongXiang Sun}, \bibinfo{person}{Xiaoxue Zang}, \bibinfo{person}{Jing Lu}, \bibinfo{person}{Yiqun Hui}, \bibinfo{person}{Xingchao Cao}, \bibinfo{person}{Zeyu Yang}, \bibinfo{person}{Yichen Zheng}, \bibinfo{person}{Dewei Leng}, {et~al\mbox{.}}} \bibinfo{year}{2024}\natexlab{}.
\newblock \showarticletitle{TWIN V2: Scaling Ultra-Long User Behavior Sequence Modeling for Enhanced CTR Prediction at Kuaishou}.
\newblock \bibinfo{journal}{\emph{Arxiv}} (\bibinfo{year}{2024}).
\newblock


\bibitem[Tang et~al\mbox{.}(2020)]%
        {ple}
\bibfield{author}{\bibinfo{person}{Hongyan Tang}, \bibinfo{person}{Junning Liu}, \bibinfo{person}{Ming Zhao}, {and} \bibinfo{person}{Xudong Gong}.} \bibinfo{year}{2020}\natexlab{}.
\newblock \showarticletitle{Progressive Layered Extraction (PLE): A Novel Multi-Task Learning (MTL) Model for Personalized Recommendations}. In \bibinfo{booktitle}{\emph{ACM Conference on Recommender Systems (RecSys)}}.
\newblock


\bibitem[Tang et~al\mbox{.}(2023)]%
        {counterfactual}
\bibfield{author}{\bibinfo{person}{Shisong Tang}, \bibinfo{person}{Qing Li}, \bibinfo{person}{Dingmin Wang}, \bibinfo{person}{Ci Gao}, \bibinfo{person}{Wentao Xiao}, \bibinfo{person}{Dan Zhao}, \bibinfo{person}{Yong Jiang}, \bibinfo{person}{Qian Ma}, {and} \bibinfo{person}{Aoyang Zhang}.} \bibinfo{year}{2023}\natexlab{}.
\newblock \showarticletitle{Counterfactual video recommendation for duration debiasing}. In \bibinfo{booktitle}{\emph{ACM SIGKDD Conference on Knowledge Discovery and Data Mining (KDD)}}.
\newblock


\bibitem[Xi et~al\mbox{.}(2023)]%
        {xi2023multimodal}
\bibfield{author}{\bibinfo{person}{Dinghao Xi}, \bibinfo{person}{Liumin Tang}, \bibinfo{person}{Runyu Chen}, {and} \bibinfo{person}{Wei Xu}.} \bibinfo{year}{2023}\natexlab{}.
\newblock \showarticletitle{A multimodal time-series method for gifting prediction in live streaming platforms}.
\newblock \bibinfo{journal}{\emph{Information Processing \& Management (IPM)}} (\bibinfo{year}{2023}).
\newblock


\bibitem[Yan et~al\mbox{.}(2024)]%
        {trinity}
\bibfield{author}{\bibinfo{person}{Jing Yan}, \bibinfo{person}{Liu Jiang}, \bibinfo{person}{Jianfei Cui}, \bibinfo{person}{Zhichen Zhao}, \bibinfo{person}{Xingyan Bin}, \bibinfo{person}{Feng Zhang}, {and} \bibinfo{person}{Zuotao Liu}.} \bibinfo{year}{2024}\natexlab{}.
\newblock \showarticletitle{Trinity: Syncretizing Multi-/Long-tail/Long-term Interests All in One}.
\newblock \bibinfo{journal}{\emph{Arxiv}} (\bibinfo{year}{2024}).
\newblock


\bibitem[Zhang et~al\mbox{.}(2023)]%
        {eliverec}
\bibfield{author}{\bibinfo{person}{Yixin Zhang}, \bibinfo{person}{Yong Liu}, \bibinfo{person}{Hao Xiong}, \bibinfo{person}{Yi Liu}, \bibinfo{person}{Fuqiang Yu}, \bibinfo{person}{Wei He}, \bibinfo{person}{Yonghui Xu}, \bibinfo{person}{Lizhen Cui}, {and} \bibinfo{person}{Chunyan Miao}.} \bibinfo{year}{2023}\natexlab{}.
\newblock \showarticletitle{Cross-domain disentangled learning for e-commerce live streaming recommendation}. In \bibinfo{booktitle}{\emph{IEEE International Conference on Data Engineering (ICDE)}}.
\newblock


\bibitem[Zhang and Yang(2022)]%
        {MTLSurvey}
\bibfield{author}{\bibinfo{person}{Yu Zhang} {and} \bibinfo{person}{Qiang Yang}.} \bibinfo{year}{2022}\natexlab{}.
\newblock \showarticletitle{A Survey on Multi-Task Learning}.
\newblock \bibinfo{journal}{\emph{IEEE Transactions on Knowledge and Data Engineering (TKDE)}} (\bibinfo{year}{2022}).
\newblock


\bibitem[Zhao et~al\mbox{.}(2019)]%
        {youtubemmoe}
\bibfield{author}{\bibinfo{person}{Zhe Zhao}, \bibinfo{person}{Lichan Hong}, \bibinfo{person}{Li Wei}, \bibinfo{person}{Jilin Chen}, \bibinfo{person}{Aniruddh Nath}, \bibinfo{person}{Shawn Andrews}, \bibinfo{person}{Aditee Kumthekar}, \bibinfo{person}{Maheswaran Sathiamoorthy}, \bibinfo{person}{Xinyang Yi}, {and} \bibinfo{person}{Ed Chi}.} \bibinfo{year}{2019}\natexlab{}.
\newblock \showarticletitle{Recommending what video to watch next: a multitask ranking system}. In \bibinfo{booktitle}{\emph{ACM Conference on Recommender Systems (RecSys)}}.
\newblock


\bibitem[Zheng et~al\mbox{.}(2023)]%
        {zheng2023dual}
\bibfield{author}{\bibinfo{person}{Jiawei Zheng}, \bibinfo{person}{Hao Gu}, \bibinfo{person}{Chonggang Song}, \bibinfo{person}{Dandan Lin}, \bibinfo{person}{Lingling Yi}, {and} \bibinfo{person}{Chuan Chen}.} \bibinfo{year}{2023}\natexlab{}.
\newblock \showarticletitle{Dual Interests-Aligned Graph Auto-Encoders for Cross-domain Recommendation in WeChat}. In \bibinfo{booktitle}{\emph{ACM International Conference on Information and Knowledge Management (CIKM)}}.
\newblock


\bibitem[Zhou et~al\mbox{.}(2019)]%
        {dien}
\bibfield{author}{\bibinfo{person}{Guorui Zhou}, \bibinfo{person}{Na Mou}, \bibinfo{person}{Ying Fan}, \bibinfo{person}{Qi Pi}, \bibinfo{person}{Weijie Bian}, \bibinfo{person}{Chang Zhou}, \bibinfo{person}{Xiaoqiang Zhu}, {and} \bibinfo{person}{Kun Gai}.} \bibinfo{year}{2019}\natexlab{}.
\newblock \showarticletitle{Deep interest evolution network for click-through rate prediction}. In \bibinfo{booktitle}{\emph{AAAI Conference on Artificial Intelligence (AAAI)}}.
\newblock


\bibitem[Zhou et~al\mbox{.}(2018)]%
        {din}
\bibfield{author}{\bibinfo{person}{Guorui Zhou}, \bibinfo{person}{Xiaoqiang Zhu}, \bibinfo{person}{Chenru Song}, \bibinfo{person}{Ying Fan}, \bibinfo{person}{Han Zhu}, \bibinfo{person}{Xiao Ma}, \bibinfo{person}{Yanghui Yan}, \bibinfo{person}{Junqi Jin}, \bibinfo{person}{Han Li}, {and} \bibinfo{person}{Kun Gai}.} \bibinfo{year}{2018}\natexlab{}.
\newblock \showarticletitle{Deep Interest Network for Click-Through Rate Prediction}. In \bibinfo{booktitle}{\emph{ACM SIGKDD Conference on Knowledge Discovery and Data Mining (KDD)}}.
\newblock


\bibitem[Zhu et~al\mbox{.}(2021)]%
        {fuxictr}
\bibfield{author}{\bibinfo{person}{Jieming Zhu}, \bibinfo{person}{Jinyang Liu}, \bibinfo{person}{Shuai Yang}, \bibinfo{person}{Qi Zhang}, {and} \bibinfo{person}{Xiuqiang He}.} \bibinfo{year}{2021}\natexlab{}.
\newblock \showarticletitle{Open benchmarking for click-through rate prediction}. In \bibinfo{booktitle}{\emph{ACM International Conference on Information and Knowledge Management (CIKM)}}.
\newblock


\bibitem[Zhu et~al\mbox{.}(2024)]%
        {interestclock}
\bibfield{author}{\bibinfo{person}{Yongchun Zhu}, \bibinfo{person}{Jingwu Chen}, \bibinfo{person}{Ling Chen}, \bibinfo{person}{Yitan Li}, \bibinfo{person}{Feng Zhang}, {and} \bibinfo{person}{Zuotao Liu}.} \bibinfo{year}{2024}\natexlab{}.
\newblock \showarticletitle{Interest Clock: Time Perception in Real-Time Streaming Recommendation System}. In \bibinfo{booktitle}{\emph{International ACM SIGIR Conference on Research and Development in Information Retrieval (SIGIR)}}.
\newblock


\end{thebibliography}
\end{document}